\documentclass[conference]{IEEEtran}
\IEEEoverridecommandlockouts
\usepackage{cite}
\usepackage{amsmath,amssymb,amsfonts}
\usepackage{algorithmic}
\usepackage{graphicx}
\usepackage{textcomp}
\usepackage{xcolor}
\def\BibTeX{{\rm B\kern-.05em{\sc i\kern-.025em b}\kern-.08em
    T\kern-.1667em\lower.7ex\hbox{E}\kern-.125emX}}

\usepackage{url}
\usepackage{enumitem}
\usepackage{caption}
\usepackage{subcaption}
\newcommand{\citet}[1]{\cite{#1}}
\renewcommand{\baselinestretch}{0.96}
\begin{document}

\title{A multi-objective time series analysis of community mobility reduction comparing first and second COVID-19 waves}

\author{\IEEEauthorblockN{Gabriela Cavalcante da Silva}
\IEEEauthorblockA{\textit{IMD, Universidade Federal} \\ 
\textit{do Rio Grande do Norte}\\
Natal, RN, Brazil \\
gabrielacavalcante@ufrn.edu.br}
\and
\IEEEauthorblockN{Fernanda Monteiro de Almeida}
\IEEEauthorblockA{\textit{PPgTI, Universidade Federal} \\ 
\textit{do Rio Grande do Norte}\\
Natal, RN, Brazil \\
feemonteiro@ufrn.edu.br}
\and
\IEEEauthorblockN{Sabrina Oliveira}
\IEEEauthorblockA{\textit{PPGMMC, Centro Federal de Educação} \\
\textit{Tecnológica de Minas Gerais}\\
Belo Horizonte, MG, Brazil \\
oliveira.sabrina@gmail.com}
\and
\IEEEauthorblockN{Leonardo C. T. Bezerra}
\IEEEauthorblockA{\textit{IMD, Universidade Federal} \\ 
\textit{do Rio Grande do Norte}\\
Natal, RN, Brazil \\
leobezerra@imd.ufrn.br}
\and
\IEEEauthorblockN{Elizabeth F. Wanner}
\IEEEauthorblockA{\textit{DECOM, Centro Federal de Educação} \\
\textit{Tecnológica de Minas Gerais}\\
Belo Horizonte, MG, Brazil \\
efwanner@cefetmg.br}
\and
\IEEEauthorblockN{Ricardo H. C. Takahashi}
\IEEEauthorblockA{\textit{DEMAT, Universidade Federal } \\
\textit{de Minas Gerais}\\
Belo Horizonte, MG, Brazil \\
taka@mat.ufmg.br}
}

% \author{
%     \IEEEauthorblockN{
%         Gabriela Cavalcante~da~Silva\IEEEauthorrefmark{1}, 
%         Fernanda Monteiro~de~Almeida\IEEEauthorrefmark{1}, 
%         Sabrina Oliveira\IEEEauthorrefmark{2},
%         Elizabeth F. Wanner\IEEEauthorrefmark{2},\\ 
%         Leonardo~C.~T. Bezerra\IEEEauthorrefmark{1} and 
%         Ricardo H. C. Takahashi\IEEEauthorrefmark{3}
%     }
%     \IEEEauthorblockA{
%         \IEEEauthorrefmark{1}IMD, Universidade Federal do Rio Grande do Norte, Natal, RN, Brazil\\
%         Email: \{gabi@,feemonteiro@,leobezerra@imd.\}ufrn.br
%     }
%     \IEEEauthorblockA{
%         \IEEEauthorrefmark{2}DECOM, Centro Federal de Educação Tecnológica de Minas Gerais, Belo Horizonte, MG, Brazil\\
%         Email: \{soliveira,efwanner\}@cefetmg.br
%     }
%     \IEEEauthorblockA{
%         \IEEEauthorrefmark{3}DEMAT, Universidade Federal de Minas Gerais, Belo Horizonte, MG, Brazil\\
%         Email: taka@mat.ufmg.br
%     }
% }

\maketitle

\begin{abstract}
With the logistic challenges faced by most countries for the production, distribution, and application of vaccines for the novel coronavirus disease~(COVID-19), social distancing~(SD) remains the most tangible approach to mitigate the spread of the virus. To assist SD monitoring, several tech companies have made publicly available anonymized mobility data. In this work, we conduct a multi-objective mobility reduction rate comparison between the first and second COVID-19 waves in several localities from America and Europe using Google community mobility reports~(CMR) data. Through multi-dimensional visualization, we are able to compare in a Pareto-compliant way the reduction in mobility from the different lockdown periods for each locality selected, simultaneously considering all place categories provided in CMR. In addition, our analysis comprises a 56-day lockdown period for each locality and COVID-19 wave, which we analyze both as 56-day periods and as 14-day consecutive windows. Results vary considerably as a function of the locality considered, particularly when the temporal evolution of the mobility reduction is considered. We thus discuss each locality individually, relating social distancing measures and the reduction observed. 
% abut in general the lockdowns enforced in the first wave were initially more effective than the lockdowns enforced in the second wave. Only after a few weeks into the second wave countries strengthened lockdowns, causing reduction rates to become higher than during the first wave.  
\end{abstract}

% \begin{IEEEkeywords}
% component, formatting, style, styling, insert
% \end{IEEEkeywords}

\section{Introduction}
\label{sec:intro}

In December 2019, the World Health Organization~(WHO) Country Office in the People’s Republic of China reported cases of pneumonia of unknown etiology. %~\cite{WHO-timeline}. 
%\footnote{\url{https://who.int/news-room/detail/29-06-2020-covidtimeline}}
In January 2020, WHO named \textit{SARS-CoV-2} the novel coronavirus responsible for these cases, and the acute respiratory syndrome it caused \text{COVID-19}. %~(\textit{coronavirus disease 2019}).
Still in January, WHO classified COVID-19 as a public health emergency of international concern. By March 2020, COVID-19 had cases reported from all continents, and WHO declared it a pandemic. Up to February 21th, 2021, a fresh worldwide figure of over 110 million % 108,153,741 
positive cases and nearly 2,5 million %2,381,295 
death records signify the severity of this viral infection, according to WHO.

% The rapid spread of COVID-19 and the deaths it produced triggered a global scientific rush. Indeed, by April 2020 over 20 surveys on COVID-19 could already be identified~\cite{Yu2020assessment}. However, this fast publication pace raises concerns on the quality of the works produced. Specifically, many such works have not undergone a peer reviewing process, casting uncertainty as to the methodological decisions adopted by their proposers. This is even more important in the COVID-19 context, where knowledge from complementary research fields is required to propose multi-disciplinary solutions to fight the pandemic.

Among the most relevant topics in multi-disciplinary COVID-19 research is social distancing~(SD), which WHO actively promotes as a non-pharmaceutical intervention against COVID-19~\cite{WHO-sd}.
%\footnote{\url{https://www.who.int/emergencies/diseases/novel-coronavirus-2019/advice-for-public}}
As the numbers of infected people and deaths increase worldwide, and while the vaccine is not globally available, health authorities have continued to rely on SD measures 
% (in combination with other everyday preventive actions) 
to reduce the spread of the disease. In fact, the lockdown policies enforced by some affected countries in 2020 helped bend the curve of the disease spread, leading to an end of the first wave. 

Despite the warnings about the consequences of premature lifting of SD measures \cite{Beware2020}, most countries eased restrictions and have already been hit by a second COVID-19 wave. Indeed, many countries in Europe have been facing a second wave COVID-19 pandemic since September 2020. It is legitimate to think that countries would manage the second wave a lot better, given the lessons learned during the first wave. However the reality in such countries tells a different scenario \cite{Graichen2021}: much higher infection numbers, more patients in ICUs, and in some countries also more deaths, implying that only a hard lockdown can help control the pandemic.

To aid SD monitoring, information technology~(IT) companies have been publishing anonymized mobile device location history data, among which Google~\cite{Aktay2020cmr}. In particular, the community mobility reports (CMR) provided by Google comprise over 130 countries, some of which further detailed on a regional level. More importantly, the data for a given locality is a collection of time series for six place categories. Therefore, SD analysis based on CMR data requires theoretical approaches from the fields of time series analysis~(TSA)~\cite{Cleveland1990stl} and multi-objective optimization~(MO)~\cite{Zitzler2003performance}. Furthermore, the existence of six different place types renders this a \textit{many-objective} optimization problem~\cite{Li2015many}, a particularly challenging specialization of MO. %Indeed, Google itself released guidelines to aid this analysis~\cite{CMR}, such its awareness of the many different research fields it bridges.

% The goal of this paper is to discuss methodological approaches that can help analyze SD measures adopted by certain localities based on CMR data and compare their effectiveness during the first and second COVID-19 waves, with a special focus on MCDM. In principle, the temporal nature of this data requires techniques from time series analysis (TSA) %, from which we comment on the effects of (i)~alternatives 
% to reduce seasonality effects~\cite{Cleveland1990stl}; techniques to aggregate temporal dynamics, and different granularities for temporal discretization. We also discuss the effects of a MCDM approach that helps provide a multi-criteria perspective to this analysis, namely Pareto dominance~\cite{Zitzler2003performance}. %and (ii)~Pareto-related measures that help deepen the analysis~\cite{Goldberg1989genetic,Laumanns2002epsilon}. 
% Moreover, we adopt visualization techniques that help understand the MCDM conclusions even in this many-criteria context.

In this paper, we employ appropriate methodology to empirically compare the community mobility reduction during the first and second COVID-19 waves in localities from America and Europe. Specifically, our analysis comprises localities that experienced some of the most severe pandemic outbreaks~\cite{jhu}, namely Lombardia~(Italy), Île-de-France~(France), and Birmingham District~(United Kingdom). Furthermore, major leader localities such as Berlin~(Germany) and Toronto Division~(Canada) have also been included, for representativeness. In common, all of these cities have enforced lockdown measures around March 2020 to contain the first wave, and were forced backed into lockdown around November 2020 due to a second wave.

Our comparison is both (i)~\textit{multi-objective}, as it simultaneously considers all place categories provided, and (ii)~\textit{temporal}, as mobility reduction is progressively compared over a 56-day period, discretized as 14-day consecutive time windows. To compare mobility reduction from a given locality during the first and second wave lockdowns in a Pareto-compliant way, we employ multi-dimensional visualization through radar charts. Furthermore, to ensure that weekday seasonal effects do not affect our analysis, we process the data using seasonal-trend decomposition by loess~(STL~\cite{Cleveland1990stl}).

%Insights observed confirm the effects and importance of the methodological resources discussed in this work. Some of these insights were already expected, such as the effects of calibration and scaling on the comparison, and the need for seasonality approaches to process the original raw data. Other insights are specific to CMR assessment, such as (i)~how the nature of the different place categories interact with the experimental factors considered and (ii)~how different temporal granularities reveal contrasting dynamics among localities. More importantly, we observe how an MCDM perspective provides nuances that scalarized approaches would conceal, and assist the decision-maker both visually and when configuring tolerance levels. Concerning localities, the region-level analysis indicates that the European regions considered were able to better promote social distancing than the American ones, though a fine-grained temporal assessment indicates that this was only achieved after a few weeks of mobility restriction. %Regarding countries, a similar geographical pattern cannot be established. 

% Below, we summarize the main contributions of this paper:
% \begin{itemize}
%     \item A theoretical discussion on a methodological alternative for a multi-criteria temporal analysis of social distancing through mobility data.
%     \item Taking into account the first and second COVID-19 waves, an empirical mobility reduction comparison of some of the most relevant outbreak examples from  different continents in a region-level.
% \end{itemize}

Results vary as a function of the locality considered, and hence we discuss each locality individually. Nonetheless, in general the initial mobility reduction in lockdown periods from the first wave was higher than in lockdown periods from the second wave. 
Alarmingly, most localities presented an increase in  mobility for all categories during the December holiday shopping season, which had not been observed during Easter holidays.
%and; (iii) the contrast-ing results forParksandGrocery & pharmacy.
%It is only after one or two 14-day time windows that the second wave lockdowns get strengthened, and mobility drops accordingly. 
Though striking, these findings are explained by the reluctance from societies and governments in general to adhere to a second lockdown, specially after the significant economical cost of the first lockdown periods.

The remainder of this paper is organized as follows. We initially briefly review related work in Section~\ref{sec:related-work}. Next, Section~\ref{sec:methodology} discusses the methodology we adopt in this work to enable a multi-objective social distancing analysis comparison from mobility data. The assessment contrasting mobility reduction from different lockdown periods for each locality is detailed in Section~\ref{sec:results}. Finally, we conclude and discuss future work in Section~\ref{sec:conclusions}.

\section{Related work}
\label{sec:related-work}

%As previously discussed, 
The research on COVID-19 is marked by (i)~the \textit{volume} of works from the most diverse scientific fields, and in particular how these fields are bridged to produced relevant multi-disciplinary insights, and; (ii)~the \textit{speed} with which works are being made available to society, though to a very large extent as pre-prints that have not yet undergone peer review. In this section, we briefly discuss works that have already been peer-reviewed. In addition, since our work compares first and second waves data, we focus on works that concern the second wave. We group these works into the most recurring topics.%, namely (i)~planning, (ii)~analysis, and (iii)~modelling. 

\begin{figure*}[!ht]
\includegraphics[width=\textwidth, clip=true, trim=10px 15px 16px 10px]{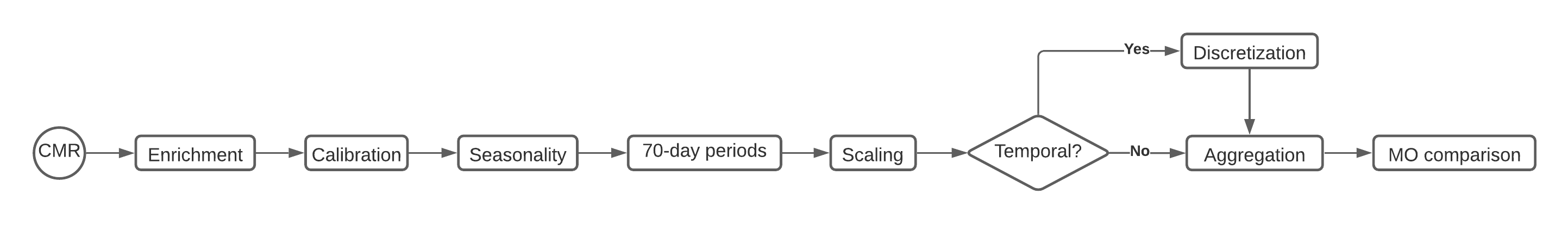}
\caption{Methodology comprising data enrichment and preparation through time series analysis and multi-objective optimization.}
\label{FIG:methodology}
\end{figure*}

\begin{description}[style=unboxed, leftmargin=0px]
\item [Planning] for a second wave has been the topic of works dating as of March 2020~\cite{LeuWuLiuLeu2020china,Beware2020,Middleton2020,MogAbdBiz2020cambridge,PanKerStuMisKleVinBon2020school,WilBatFis2020travel}. Given how early China was able to contain COVID-19 spread during its first wave, \cite{LeuWuLiuLeu2020china} discuss the benefits of social distancing outside Hubei and stress the importance of second wave planning. Building on that work, \cite{Beware2020} state the need to model how different non-pharmaceutical interventions such as SD individually contributed to contain or mitigate the epidemic in China. In particular, authors emphasize understanding these individual contributions as key to second wave prevention planning.

Outside China, where first waves were not under control until spring 2020, %
%\footnote{Considering a northern hemisphere season calendar.}
second wave planning works can be identified as of July 2020~\cite{WilBatFis2020travel,Middleton2020,MogAbdBiz2020cambridge,PanKerStuMisKleVinBon2020school}. Building on the lessons from China, social distancing is further advocated for by~\citet{WilBatFis2020travel}. In~\cite{Middleton2020}, authors discuss how the second wave was expected to hit Europe harder than the first wave, and highlight the effects of the winter season in northern hemisphere countries. \cite{MogAbdBiz2020cambridge} stress that until herd immunity has been achieved, not only a second wave but multiple peaks in COVID-19 spread should be expected. Finally, besides traditional non-pharmaceutical approaches discussed in other works, \cite{PanKerStuMisKleVinBon2020school} also include school reopening as a planning factor for a second wave.

\item [Analysis] works date of after the second waves had started across Europe~\cite{GreCus2020attenuated,SaiAsaMatHayTerOhtTsuOhm2020japan,DiaVer2020brazil,ConFraPajTirMenPla2021france,Graichen2021,BONTEMPI2021,IoaAxfCon2021shift}, and focus on the development and characteristics of the second waves.  Works from this category are as early as fall 2020, but the first ones assessed the second wave at a point where daily deaths were attenuated in comparison to the first wave~\cite{GreCus2020attenuated,SaiAsaMatHayTerOhtTsuOhm2020japan}. In particular, \cite{GreCus2020attenuated} performed this comparison globally, whereas \cite{SaiAsaMatHayTerOhtTsuOhm2020japan} focused on COVID-19 waves in Japan. 

By winter 2020, the rise in the number of fatalities demonstrated that the second wave was harder than the first one~\cite{DiaVer2020brazil,ConFraPajTirMenPla2021france,Graichen2021,BONTEMPI2021}. Globally, \cite{DiaVer2020brazil} discuss the role of age and reinfection as distinguishing characteristics of the second wave. Other works focus on European nation-wide realities, namely France~\cite{ConFraPajTirMenPla2021france}, Germany~\cite{Graichen2021}, and Italy~\cite{BONTEMPI2021}. More recently, \cite{IoaAxfCon2021shift} assessed shifts in age distribution and nursing home fatalities in 16 representative European Union countries, arguiging that first and second waves are similar concerning the former, but the latter was reduced during the second wave.

\item [Modelling]
is the focus of the remaining works we identify~\cite{secwave_playbook,Second_wave_SEIR, AMCPPal2020,Vaidal2020,Renardy2020}. 
%The second wave of COVID-19 across Europe has been predicted in \cite{secwave_playbook}. 
In more detail, some of these works focus on Europe~\cite{secwave_playbook,Second_wave_SEIR}, whereas others target America~\cite{AMCPPal2020,Vaidal2020,Renardy2020}. Regarding Europe, \cite{secwave_playbook} provide projections of temporal evolution of COVID-19 spread across different regions calibrated on first wave data, and model the impact of the corresponding SD. In \cite{Second_wave_SEIR}, authors model COVID-19 second wave infections in France and Italy via a stochastic susceptible-exposed-infected-recovered model. 

Regarding America, in \cite{AMCPPal2020} authors build an agent-based model of SARS-CoV-2
transmission in the Boston metropolitan area using anonymized, geolocalized mobility data with census, and demographic data. 
%severe acute respiratory syndrome coronavirus 2 
In particular, they show that hard SD measures coupled with testing, contact-tracing, and household quarantine could not only to prevent the failure of the healthcare system but also allow the return of economic activities.
In \cite{Vaidal2020}, a multi-disciplinary approach employs a Bayesian susceptible-infected-recovered model, Kalman Filter, and machine learning techniques to investigate the effects from SD policies in North America. In \cite{Renardy2020}, considering as factors casual and workplace contacts as well as reopening speed, authors predict the most prudent action for controlling the second wave of COVID-19 in Michigan, US.
%Washtenaw County, MI, US. 

\end{description}

As discussed, the volume and speed with which scientific works on COVID-19 are being published is overwhelming. 
% Our work fits the second category, namely second wave analysis. 
Nonetheless, we have not found peer-reviewed works comparing SD measures adopted by localities during the first and second waves. In the next section, we describe the methodology we propose for such an analysis.

\section{Methodology}
\label{sec:methodology}

To compare social distancing in the first and second COVID-19 waves, we assess the mobility data provided as community mobility reports~(CMR) by Google. In this section, we detail data enrichment and preparation, including the time series analysis (TSA) and multi-objective optimization~(MO) techniques we adopt for their analysis. Figure~\ref{FIG:methodology} summarizes the process described in this section.

% In this section, we initially detail CMR data and how we enrich it with mobility restriction-related dates and prepare it for analysis. Next, we discuss time series analysis concepts and techniques, critical to this assessment as the CMR data is provided as a collection of time series. Finally, we discuss concepts and techniques from multi-criteria decision making~(MCDM) which allow us to assess the CMR data from a multi-criteria perspective.  %The whole process described in this section is summarized in Figure~\ref{FIG:methodology}.

\subsection{Data acquisition, description, and enrichment}

CMR data is provided as a comma-separated values~(CSV) file comprising over 135 countries, some of which further detailed on a regional level. Data is collected from users who willingly enable their location history, and is anonymized as described in~\citet{Aktay2020cmr}. CMR per-locality data comprises six time series, one for each place category created by Google, given in Table~\ref{tab:place_categories}. Each time series currently spans one year, having started on February 15$^\text{th}$, 2020. For a single timestamp and category, the given value is computed relatively to a baseline, namely the median value, for the corresponding day of the week, computed for the period between January 3$^\text{rd}$, 2020 and February 6$^\text{th}$, 2020~\cite{CMR}. %Since mobility for the lockdown periods is lower than for the baseline period, we refer to this relative mobility as \textit{mobility reduction rates}.

In this work, we restrict our analysis to mobility in non-residential areas. Our rationale is that the indication that residential mobility should be maximized is not as clear as that mobility in non-residential places should be minimized. More precisely, Google does not specify the space granularity it adopts for residential places. As such, cultural traits such as social gatherings from neighbors that live in a same building would have to be minimized, but Google does not disclaim how this category is processed.

% CMRs are provided by Google weekly as printed document files~(PDFs), first created on March 2020. Additionally, Google makes CMR data available as a comma-separated values~(CSV) file, which comprises data from all reports ever produced. In total, CMR data comprises 135 countries, some of which are further detailed on a regional level. Data is collected from users who willingly enable their location history, and is anonymized as described in~\citet{Aktay2020cmr}.

% Data for each country is provided as a collection of six time series, one for each place category created by Google, given in Table~\ref{tab:place_categories}. Each time series starts in February 15$^\text{th}$, 2020, and currently extends until \textcolor{red}{February 14$^\text{th}$, 2021}. For a single timestamp and category, the given value is computed relatively to a baseline, namely the median value, for the corresponding day of the week, computed for the period between January 3$^\text{rd}$, 2020 and February 6$^\text{th}$, 2020~\cite{CMR}.

\begin{table}[!t]
    \centering
    \caption{Place category descriptions from Google CMR~\cite{CMR}.}
    \label{tab:place_categories}
    \scalebox{0.8}{
    \begin{tabular}{p{2cm}|p{8cm}}
   %  \begin{tabular}{|}
        \hline
        \textbf{Category} & \textbf{Places} \\ 
        \hline
        \textit{Grocery}\linebreak \& \textit{pharmacy} & Grocery markets, food warehouses, farmers markets, specialty shops, drug stores, and pharmacies\\ 
        \hline
        \textit{Parks} & Parks, public beaches, marinas, dog parks, plazas, and public gardens\\
        \hline
        \textit{Transit stations} & Public transport hubs, e.g. subway and bus stations\\
        \hline
        \textit{Retail}\linebreak \& \textit{recreation} & Restaurants, cafes, shopping centers, theme parks, museums, libraries, and movie theaters\\
        \hline
        \textit{Residential} & Places of residence\\
        \hline
        \textit{Workplaces} & Places of work\\
        \hline
    \end{tabular}
    }
\end{table}
As a representative sample in terms of second waves, we select Lombardia (Italy), Île-de-France (France), and Birmingham District (United  Kingdom). In addition, we also include Berlin (Germany)
%
%\footnote{The data for Berlin had three missing points, which we imputed using 7-day moving average.}
and Toronto Division (Canada) to account for localities where the second wave was not as hard and/or are situated outside Europe. We remark that these localities comprise different CMR spatial discretization granularities, with data for Lombardia and Île-de-France representing the second granularity level~(a region) and data for the remaining localities representing the third granularity level~(e.g., a district or a division). Nonetheless, since we only compare data from first and second waves for each locality individually, this difference in granularity does not affect our conclusions.

\begin{figure}
    \centering
    \includegraphics[width=\linewidth]{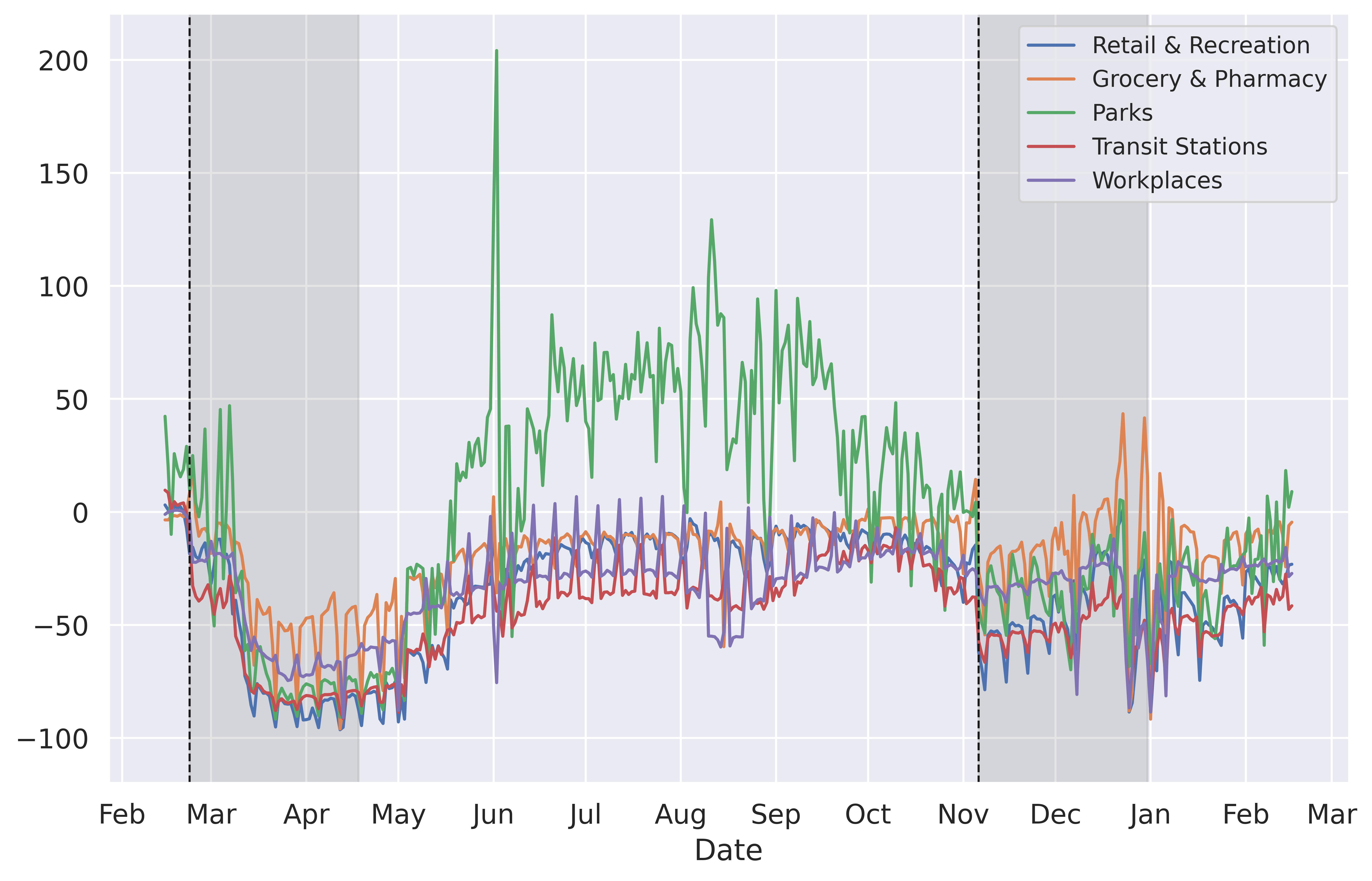}
    \caption{Original CMR data for Lombardia. Shadowed zones indicate the 56-day periods from each COVID-19 wave considered for the analysis, which start on the corresponding initial restriction date~(vertical dashed lines).}
    % \caption{Original~(left) and processed~(right) CMR data for Lombardia. Shadowed zones indicate the 56-day periods from each COVID-19 wave considered for the analysis, which start on the corresponding initial restriction date~(vertical dashed lines).}
    \label{fig:preparation}
\end{figure}

For the localities we assess, we have further enriched the data with their initial mobility restriction dates related to the first and second waves, given in Table~\ref{tab:restriction_dates}. %, and (ii)~the first wave dates when these restrictions started to be relaxed. 
Specifically, in the first wave, we have used school suspension dates as the initial restriction dates, %
%\footnote{Mid-March 2020 for all localities but Lombardia, where school suspension dates of the last week of February 2020.}
 as our preliminary assessment showed this was the restriction measure that most significantly affected mobility. In the second wave, since schools have remained open in some places, the initial restriction dates are different for each location, but not much apart from each other. %
 %\footnote{Early November for European localities, and mid-November for Toronto}.
 Nonetheless, since our assessment is based on days since initial restriction dates, differences in starting dates do not affect our conclusions. Figure~\ref{fig:preparation}~(left) illustrates CMR data for Lombardia, as well as the restriction dates for each wave for that locality~(vertical dashed lines). 
 
 \begin{table}[!t]
    \caption{Initial restriction dates for first and second waves, in 2020, considering each locality. }
    \label{tab:restriction_dates}
    \centering
    \scalebox{0.8}{
    \begin{tabular}{c|c|c|c|c|c}
        \hline
        \textbf{Wave} &  Lombardia &  Île-de-France & Birmingham & Berlin & Toronto\\
                      &          &                &
        District    &         & Division     \\
        \hline
        \textit{1st} &  Feb 23rd& Mar 12nd & Mar 13nd &  Mar 13nd &  Mar 12nd  \\ 
          %    & Final   &       &   &  & &    \\ 
        \hline
       \textit{2nd} &   Nov 6th& Oct 30th   & Nov 5th & Nov 2nd & Nov 21st   \\ 
      %        & Final   &       &   &  & &    \\ 
        
        \hline
    \end{tabular}}
\end{table}

\subsection{Data preparation}

We follow guidelines provided by Google for the assessment of CMR data~\cite{CMR}. Specifically, Google recommends (i)~calibrating data in a locality-wise basis; (ii)~handling noise incurred by holidays or other exceptional circumstances, and; (iii)~balancing the difference in magnitude between categories. 

To meet these guidelines, we first process the whole time series for each locality and place category to ensure that the data previous to the first wave first mobility restriction date present zero mean. Next, we isolate data trend from weekday seasonality effects and noise using seasonal-trend decomposition by loess~(\textbf{STL},~\cite{Cleveland1990stl}). We remark that monthly seasonality effects cannot be addressed with the currently available CMR data, as the current span of the data is only one-year long. Last, we balance the contribution of individual place categories, i.e., data for each category is scaled to a common range per locality, as done in the MO literature. 

To compare mobility reduction during the first and second waves, we focus on the 56-day periods in each wave after the corresponding initial restriction date. We select this 56-day duration for two reasons. First, the first lockdowns lasted no longer than 60 days for the localities considered, in general. In addition, a 56-day period can be further discretized into 14-day windows, matching the maximum incubation period for SARS-CoV-2. Figure~\ref{fig:preparation} depicts the 56-day periods for Lombardia~(shadowed areas) using the original data, as discussed. %By contrast, Figure~\ref{fig:lombardia5_14}~(left) depicts data after preparation. As discussed, monthly seasonal effects are still present in the prepared data, e.g., the summer holiday reduction effect on \textit{Workplaces} mobility.

To draw Pareto-compliant conclusions, we compare the prepared data from the different waves for a given locality using radar charts. Whichever the temporal granularity considered (whole 56-day period or 14-day consecutive windows), per-category data from different waves are aggregated using the area under the curve~(AUC) approach to enable comparison. 
%Figure~\ref{fig:preparation}~(right) depicts data after preparation. 
Given the use of prepared and aggregated data, the direct interpretability of mobility reduction in radar charts is lost. Yet, this approach increases the soundness of the insights we discuss in the next section.

\medskip
\section{Results}
\label{sec:results}

Social distancing~(SD) has been promoted by WHO as a critical non-pharmaceutical intervention against COVID-19, and hence localities have proposed mobility restriction measures to enforce it during each wave. Given the contrast between approaches taken by each locality, we start this section providing a unified perspective into the measures adopted. We then proceed to a per-locality discussion, where we first discuss results from a 56-day period perspective, and then from a temporal evolution perspective considering 14-day consecutive windows. 

\subsection{Social distancing measures}

Though we do not compare different localities directly, we initially discuss common SD measures adopted by localities to provide context to the analysis we conduct next. Table~\ref{tab:lockdowns} indicates whether an SD measure affecting the mobility for the given category has been enforced in the selected locality during the given COVID-19 wave. We remark that we: (i)~do not include \textit{Residential}, as it is not considered in our assessment; (ii)~do not include \textit{Grocery \& pharmacy}, as no locality has enforced measures directly related to this category, and; (iii)~include \textit{Schools} as a separate column, as school closing potentially affected multiple categories.

\begin{table*}[!t]
    \caption{Social distance restriction measures adopted by the selected localities during the first and second COVID-19 waves.}
    \label{tab:lockdowns}
    \centering
    \scalebox{0.85}{
    \begin{tabular}{c|*{5}{|c}|*{5}{|c}}
        \cline{1-11}
        & \multicolumn{5}{|c||}{\textbf{1st wave}} & \multicolumn{5}{c}{\textbf{2nd wave}}\\
        \hline
        \textbf{Locality} &  \it{Parks} & \it{Retail \& recreation} &  \it{Transit stations} & \it{Workplaces} & \it{Schools} &  \it{Parks} & \it{Retail \& recreation} &  \it{Transit stations} & \it{Workplaces} & \it{Schools}\\
        \hline
        \textit{Berlin} & $\checkmark$   & $\checkmark$  & $\checkmark$  & \textit{In part}  & $\checkmark$  & $\checkmark$  & \textit{In part}  & --- & \textit{In part}   & --- \\ 
        \hline
        \textit{Birmingham District} & ---   & $\checkmark$  & ---  & \textit{In part}  & $\checkmark$  & ---   & $\checkmark$  & --- & \textit{In part}   & --- \\ 
       
        \hline
        \textit{Île-de-France} & $\checkmark$   & $\checkmark$  & $\checkmark$  & $\checkmark$  & $\checkmark$  & \textit{In part}   & $\checkmark$  & \textit{In part} & \textit{In part}   & --- \\ 
        \hline
        \textit{Lombardia} & $\checkmark$   & $\checkmark$  & $\checkmark$  & $\checkmark$  & $\checkmark$  & \textit{In part}  & \textit{In part}  & $\checkmark$  & $\checkmark$  & --- \\ 
        \hline
        \textit{Toronto Division} & $\checkmark$   & $\checkmark$  & \textit{In part}  & \textit{In part}  & $\checkmark$  & $\checkmark$  & $\checkmark$  & \textit{In part}  & \textit{In part}  & $\checkmark$ \\ 
        \hline
    \end{tabular}}
\end{table*}

Two insights deserve highlighting at this point. First, for the first wave set of measures Île-de-France and Lombardia adopted stricter measures than the remaining European localities (especially Birmingham District). Second, comparing first and second wave restriction measures, we notice how stricter measures were during the first COVID-19 wave. As discussed, this is an effect of the socioeconomical toll incurred by the first wave, which made governments and societies less prone to restrictions when preparing for a second wave. Nonetheless, most of the localities later adopted stricter measures such as school closing as of January 2021, though it was also common to observe measures being relaxed for the December holiday shopping season.

\newcommand{\blueten}{\cellcolor{blue!10}}
\newcommand{\bluetwenty}{\cellcolor{blue!20}}
\newcommand{\bluethirty}{\cellcolor{blue!30}}
\newcommand{\bluethirtyfive}{\cellcolor{blue!35}}
\newcommand{\blueforty}{\cellcolor{blue!40}}
\newcommand{\bluefifty}{\cellcolor{blue!55}}

\newcommand\colorgd[1]{\cellcolor{blue!#10}#1}

\newcommand{\figratio}{1.0}
\newcommand{\subfigratio}{0.19}
\begin{figure*}
    \centering
    \begin{subfigure}{\subfigratio\linewidth}
    \centering
    \includegraphics[width=\figratio\linewidth, clip=true, trim=80px 45px 70px 70px]{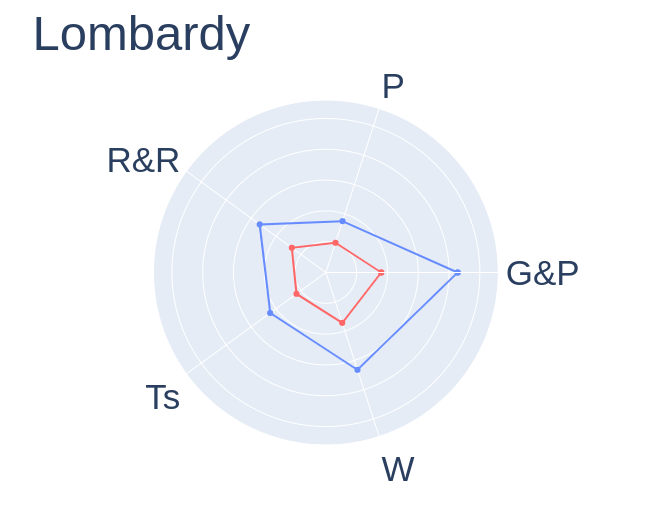}
    	\caption{ Lombardia}
		\label{FIG:lombardia70}
	\end{subfigure}
	\hfill
	 \centering
    \begin{subfigure}{\subfigratio\linewidth}
    \centering
    \includegraphics[width=\figratio\linewidth, clip=true, trim=80px 45px 70px 70px]{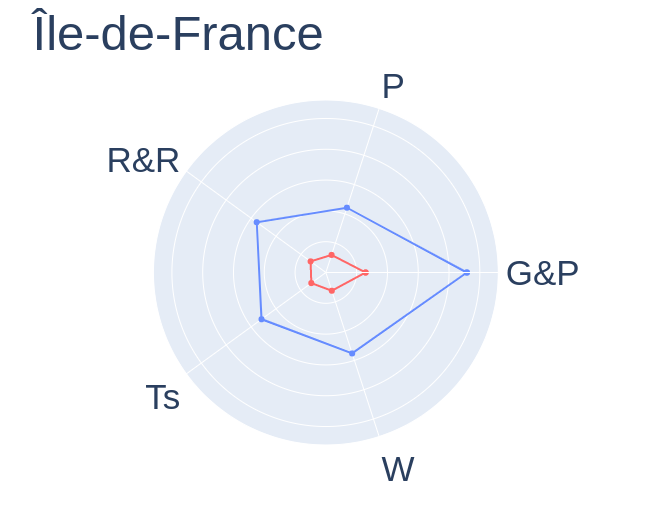}
    	\caption{ Île-de-France}
		\label{FIG:idf70}
	\end{subfigure}
	\hfill
     \centering
    \begin{subfigure}{\subfigratio\linewidth}
    \centering
    \includegraphics[width=\figratio\linewidth, clip=true, trim=80px 45px 70px 70px]{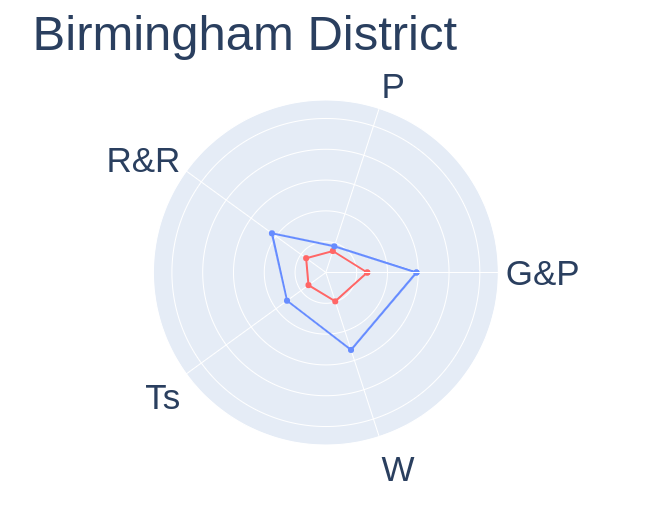}
    	\caption{ Birmingham District}
		\label{FIG:birmingham70}
	\end{subfigure}
	\hfill
     \centering
    \begin{subfigure}{\subfigratio\linewidth}
    \centering
    \includegraphics[width=\figratio\linewidth, clip=true, trim=80px 45px 70px 70px]{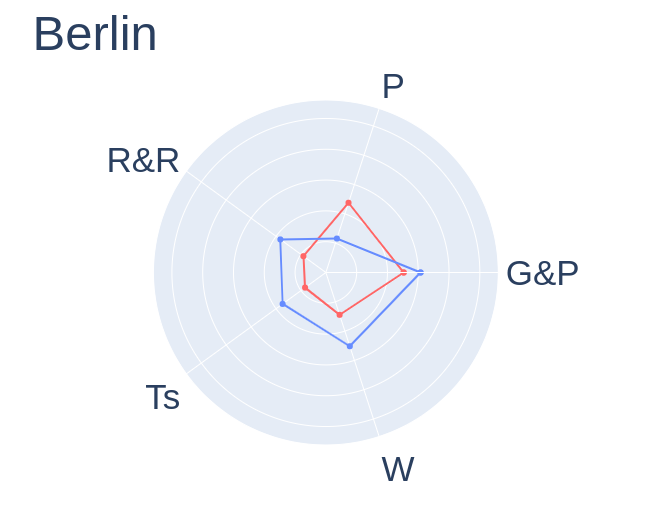}
    	\caption{ Berlin}
		\label{FIG:berlin70}
	\end{subfigure}
	\hfill
     \centering
    \begin{subfigure}{\subfigratio\linewidth}
    \centering
    \includegraphics[width=\figratio\linewidth, clip=true, trim=80px 45px 70px 70px]{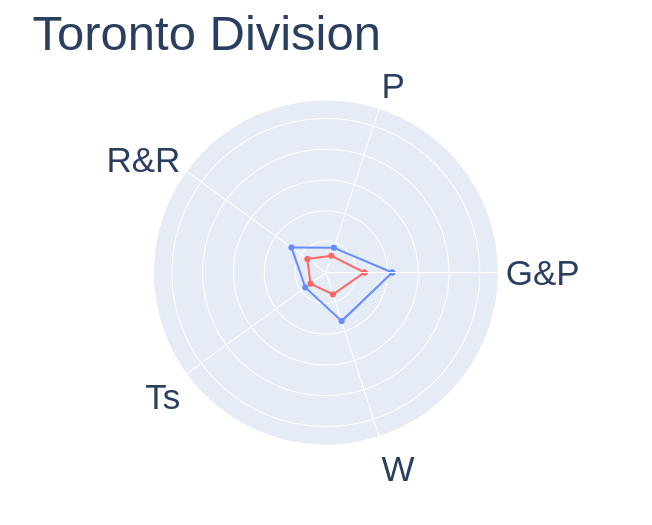}
    	\caption{ Toronto Division}
		\label{FIG:toronto70}
	\end{subfigure}
	\hfill
    \caption{Multi-objective comparison of the mobility reduction rates in the first~(red) and second~(blue) COVID-19 waves of the selected localities. Values for each (abbreviated) place category are given by the AUC aggregation of the whole 56-day period. Charts range from the minimum~(0) to the maximum~(56) possible AUC value for the period considered. %W: \textit{Workspace}; G\&P: \textit{Grocery \& Pharmacy}; P: \textit{Parks}; R \& R: \textit{Retail \& Recreation}; Ts: \textit{Transit stations}
    }
    \label{FIG:my_label}
\end{figure*}

%clip=true, trim=80px 60px 0 80px

\newcommand{\singlefigratio}{0.65}
\begin{figure*}
    \centering
    \begin{subfigure}{\linewidth}
    \centering
    \includegraphics[width=0.34\linewidth]{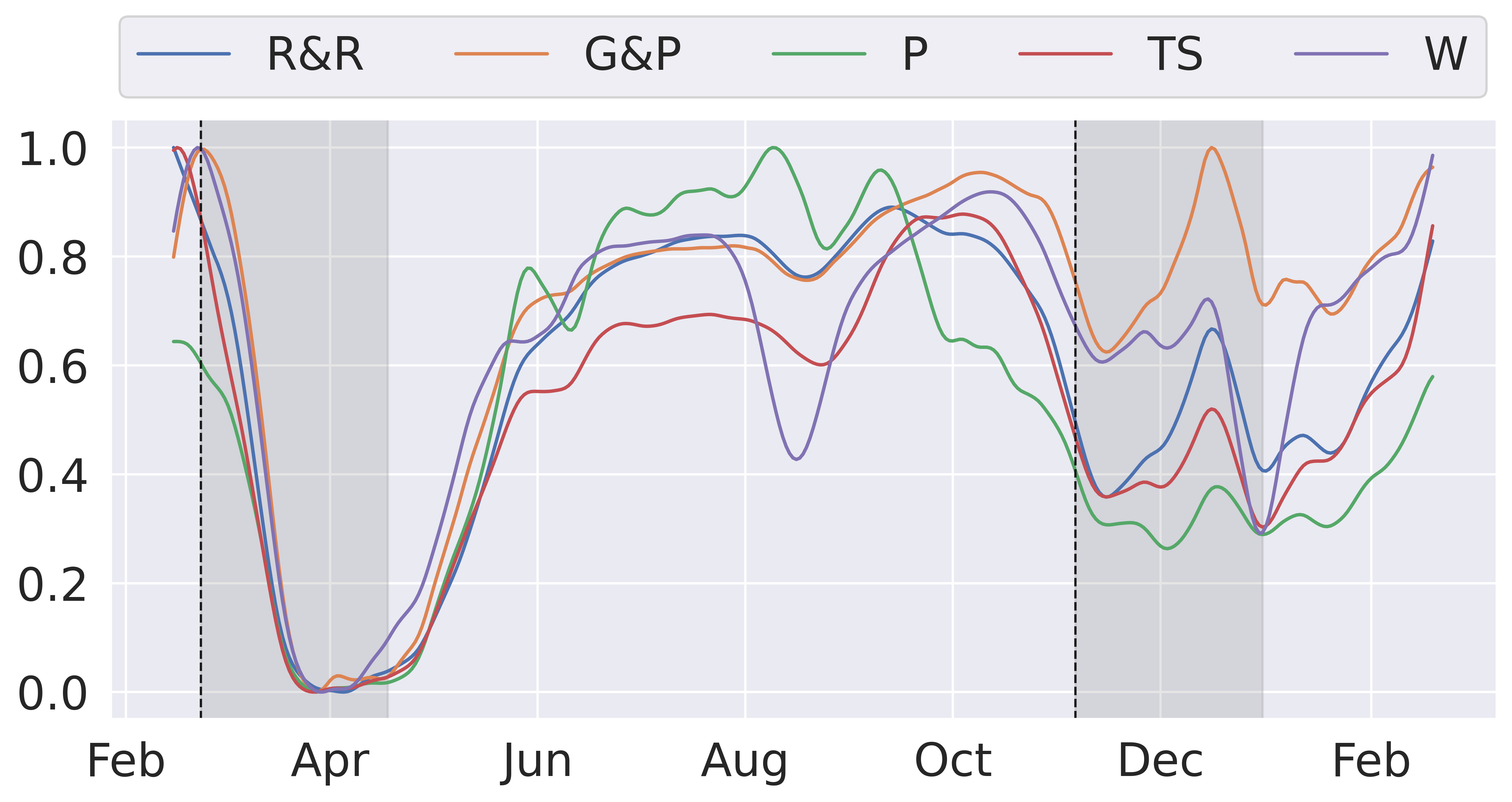}
    \includegraphics[width=\singlefigratio\linewidth, clip=true, trim=100px 30px 100px 30px]{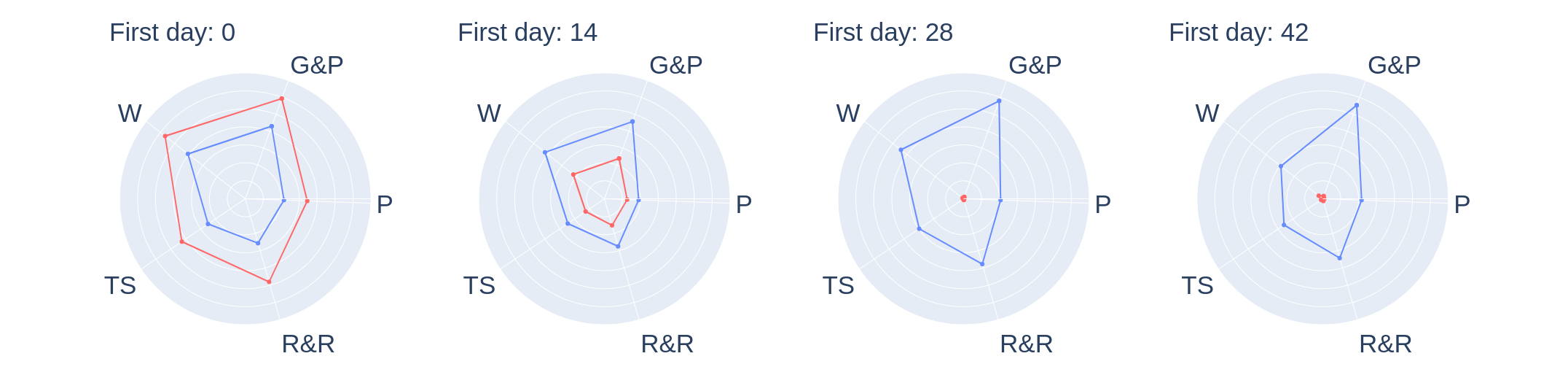}
    	\caption{ Lombardia}
		\label{fig:lombardia5_14}
	\end{subfigure}
	\hfill
	 \centering
    \begin{subfigure}{\linewidth}
    \centering
    \includegraphics[width=0.34\linewidth, clip=true, trim=0 0 0 0px]{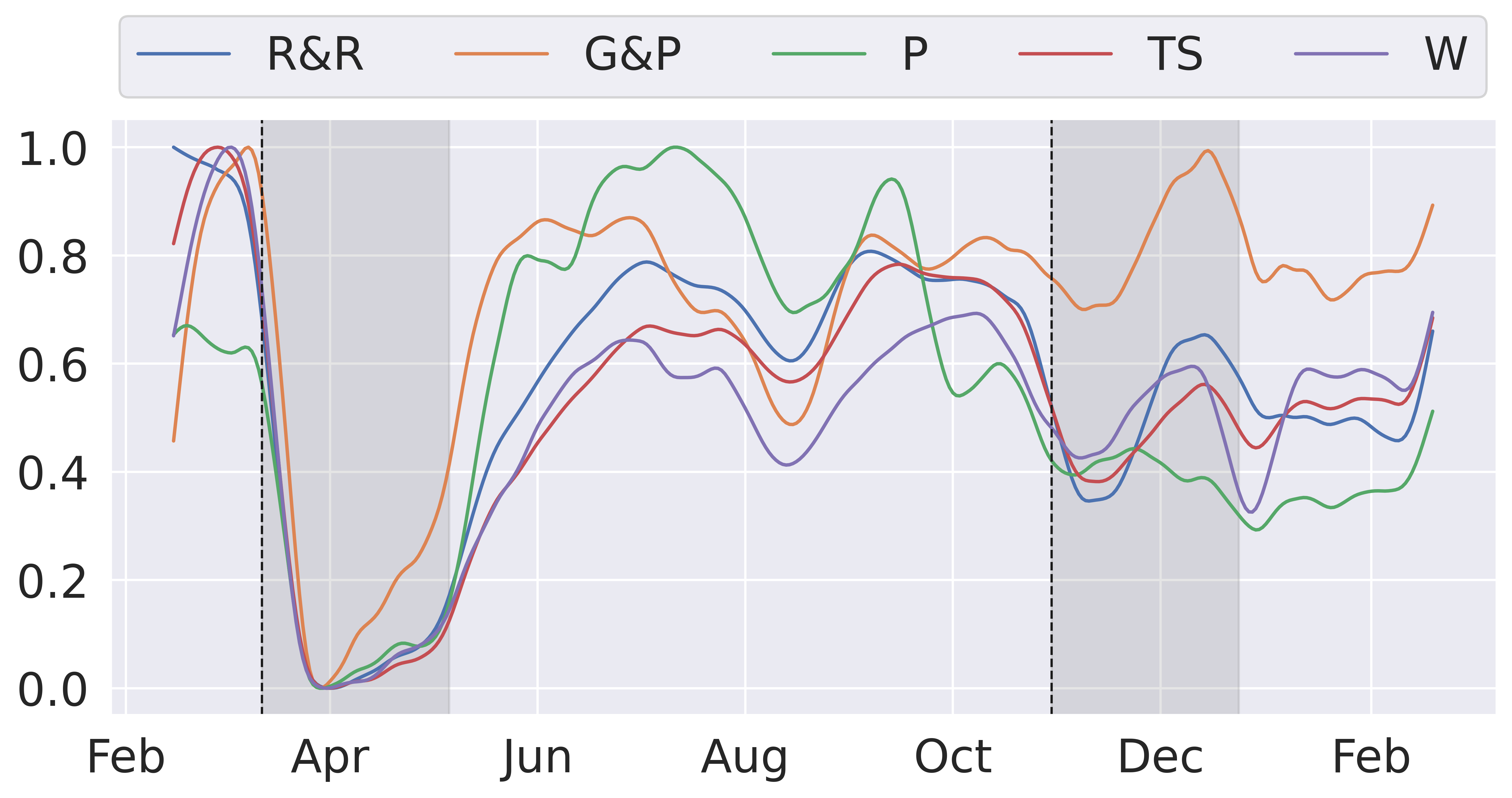}
    \includegraphics[width=\singlefigratio\linewidth, clip=true, trim=100px 30px 100px 0px]{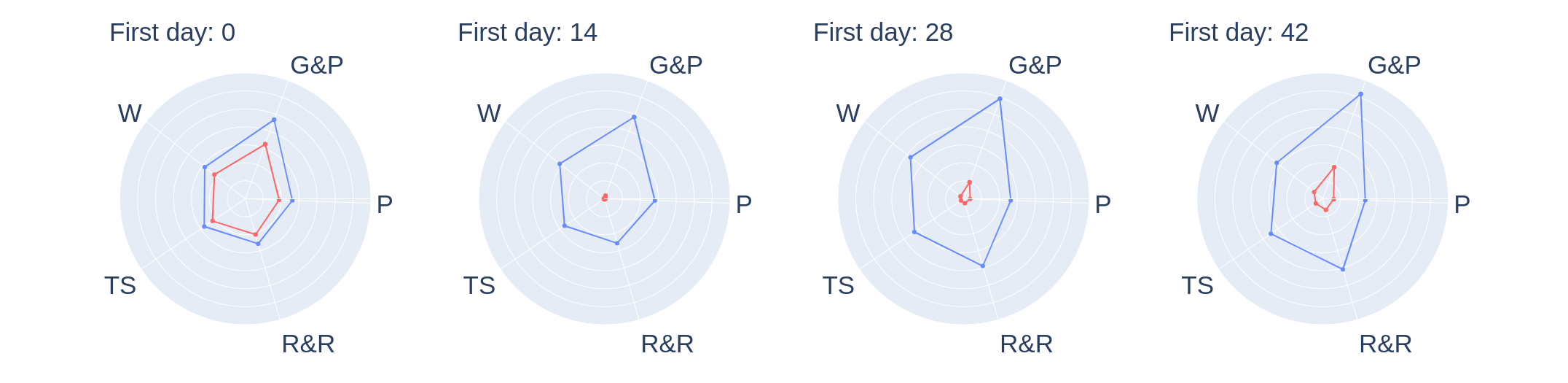}
    	\caption{ Île-de-France}
		\label{fig:idf5_14}
	\end{subfigure}
	\hfill
     \centering
    \begin{subfigure}{\linewidth}
    \centering
    \includegraphics[width=0.34\linewidth, clip=true, trim=0 0 0 0px]{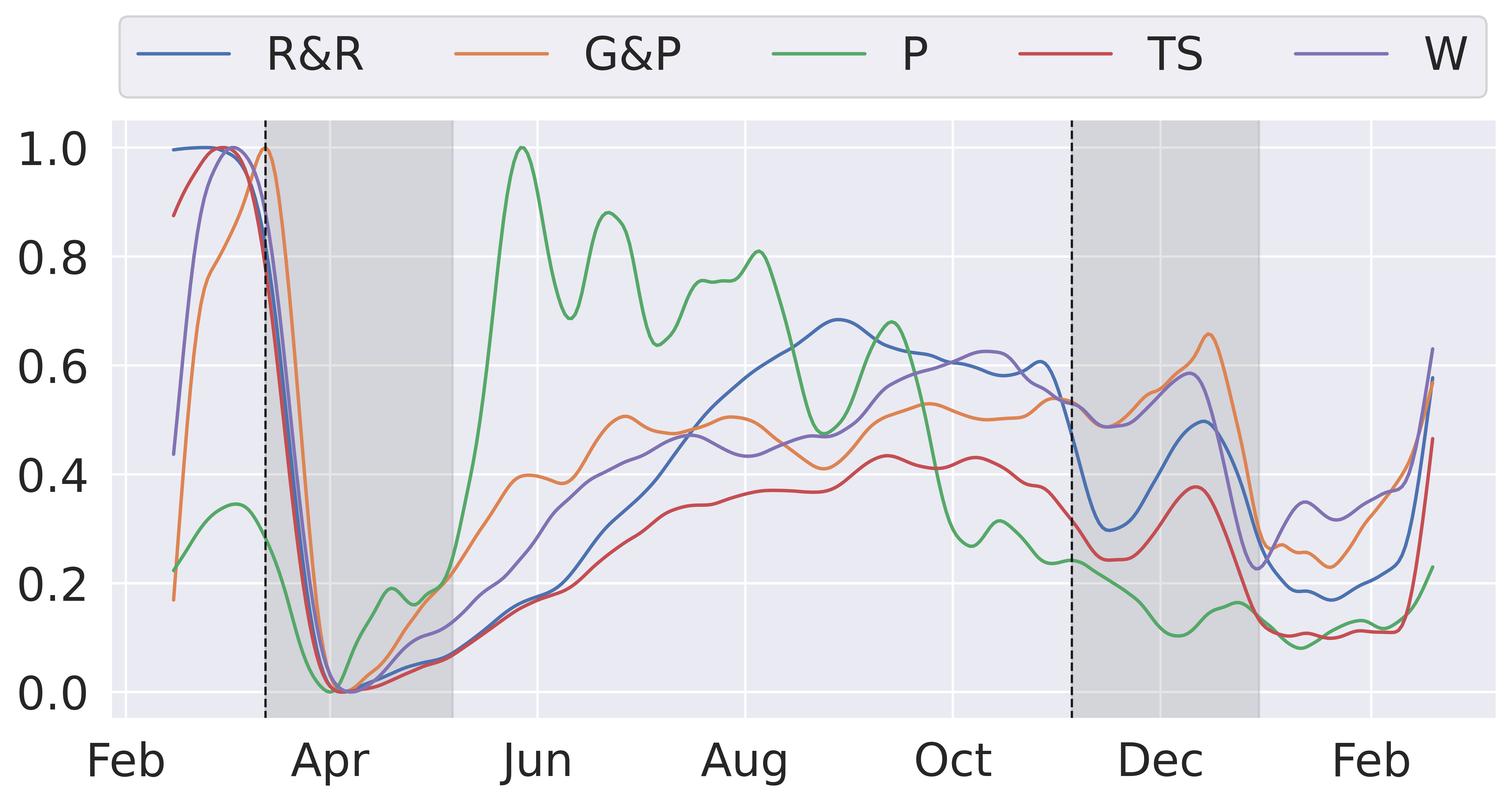}
    \includegraphics[width=\singlefigratio\linewidth, clip=true, trim=100px 30px 100px 0px]{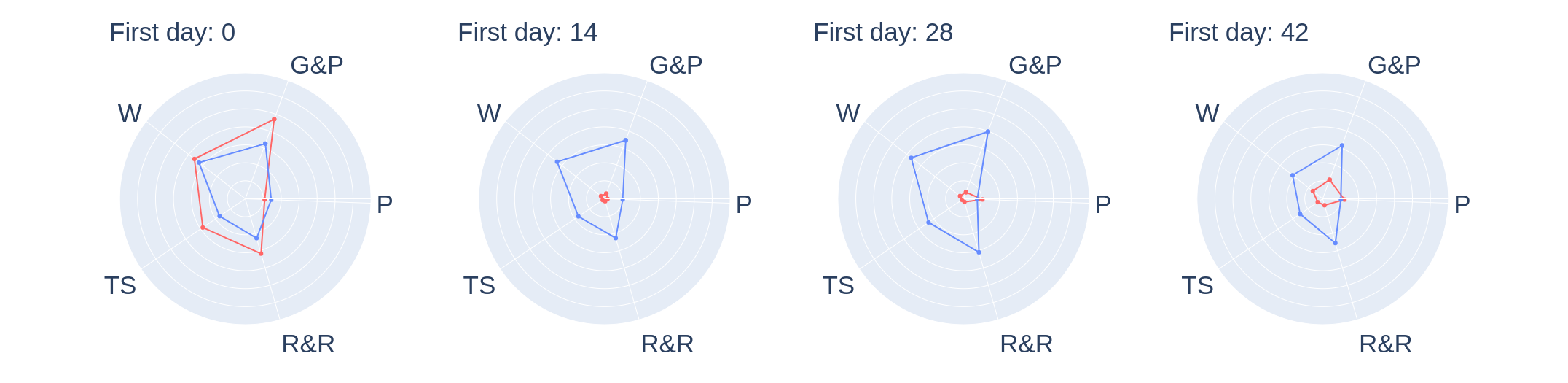}
    	\caption{ Birmingham District}
		\label{fig:birmingham5_14}
	\end{subfigure}
	\hfill
     \centering
    \begin{subfigure}{\linewidth}
    \centering
    \includegraphics[width=0.34\linewidth, clip=true, trim=0 0 0 0px]{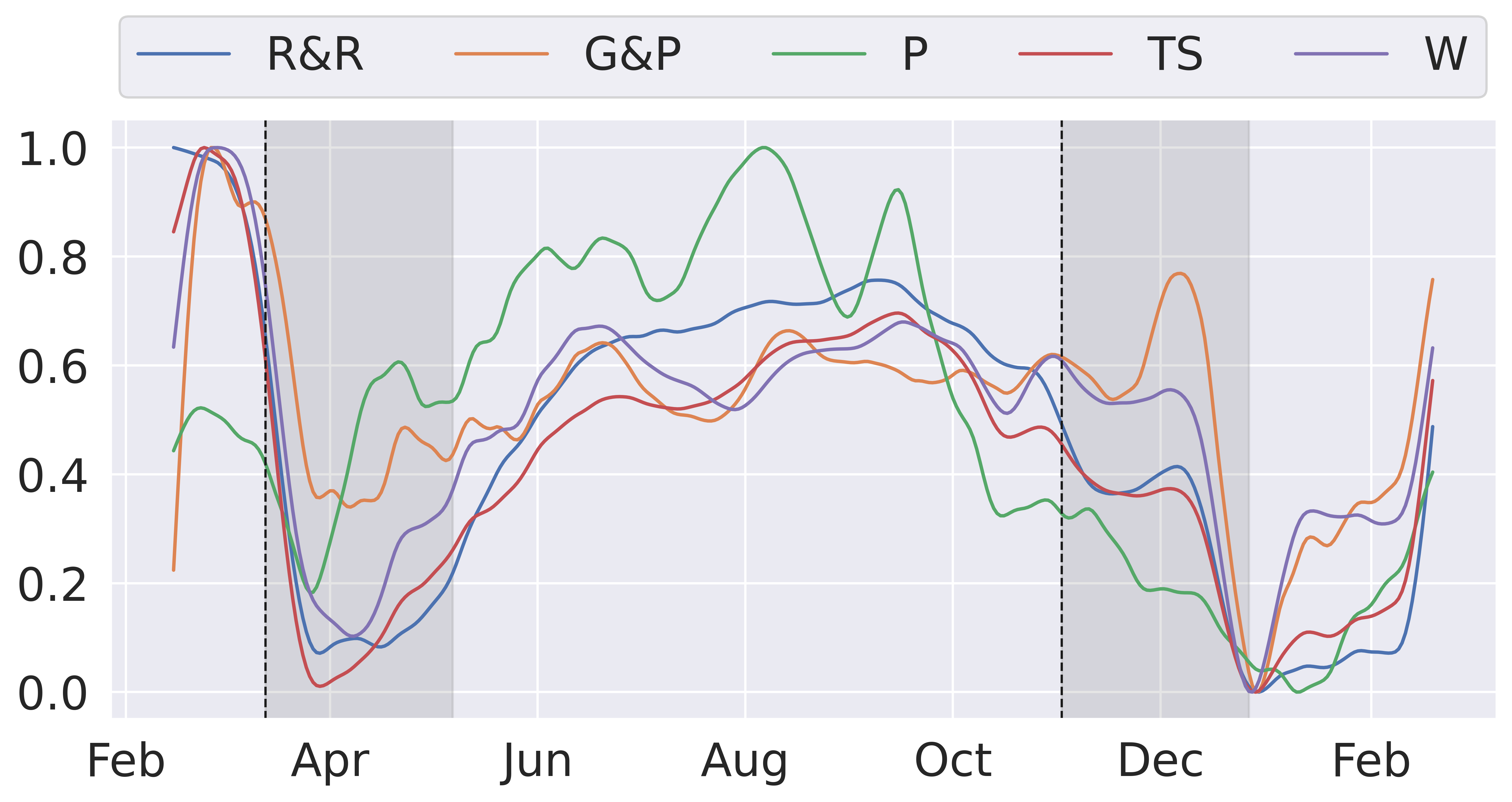}
    \includegraphics[width=\singlefigratio\linewidth, clip=true, trim=100px 30px 100px 0px]{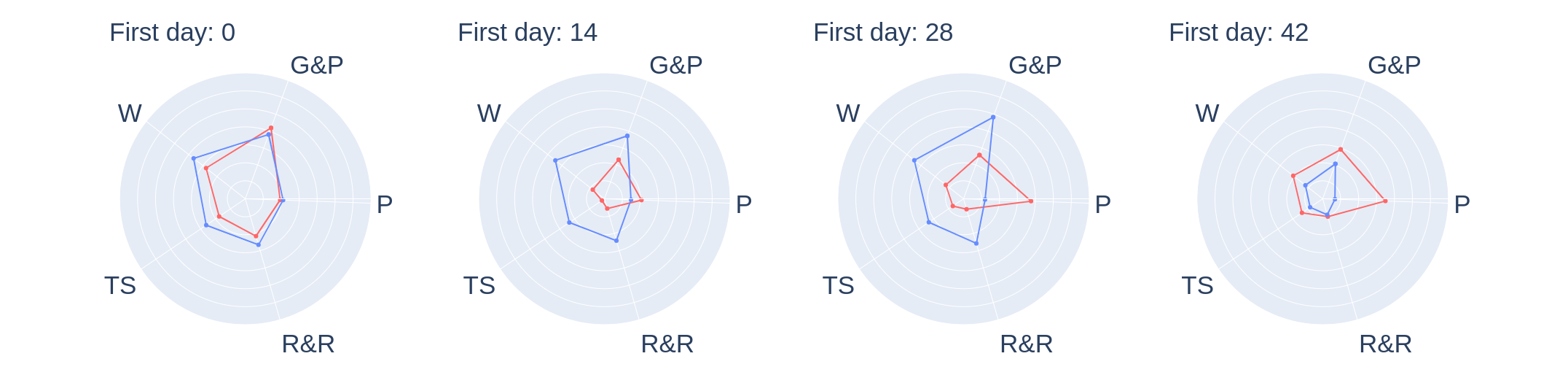}
    	\caption{ Berlin}
		\label{fig:berlin5_14}
	\end{subfigure}
	\hfill
     \centering
    \begin{subfigure}{\linewidth}
    \centering
    \includegraphics[width=0.34\linewidth, clip=true, trim=0 0 0 0px]{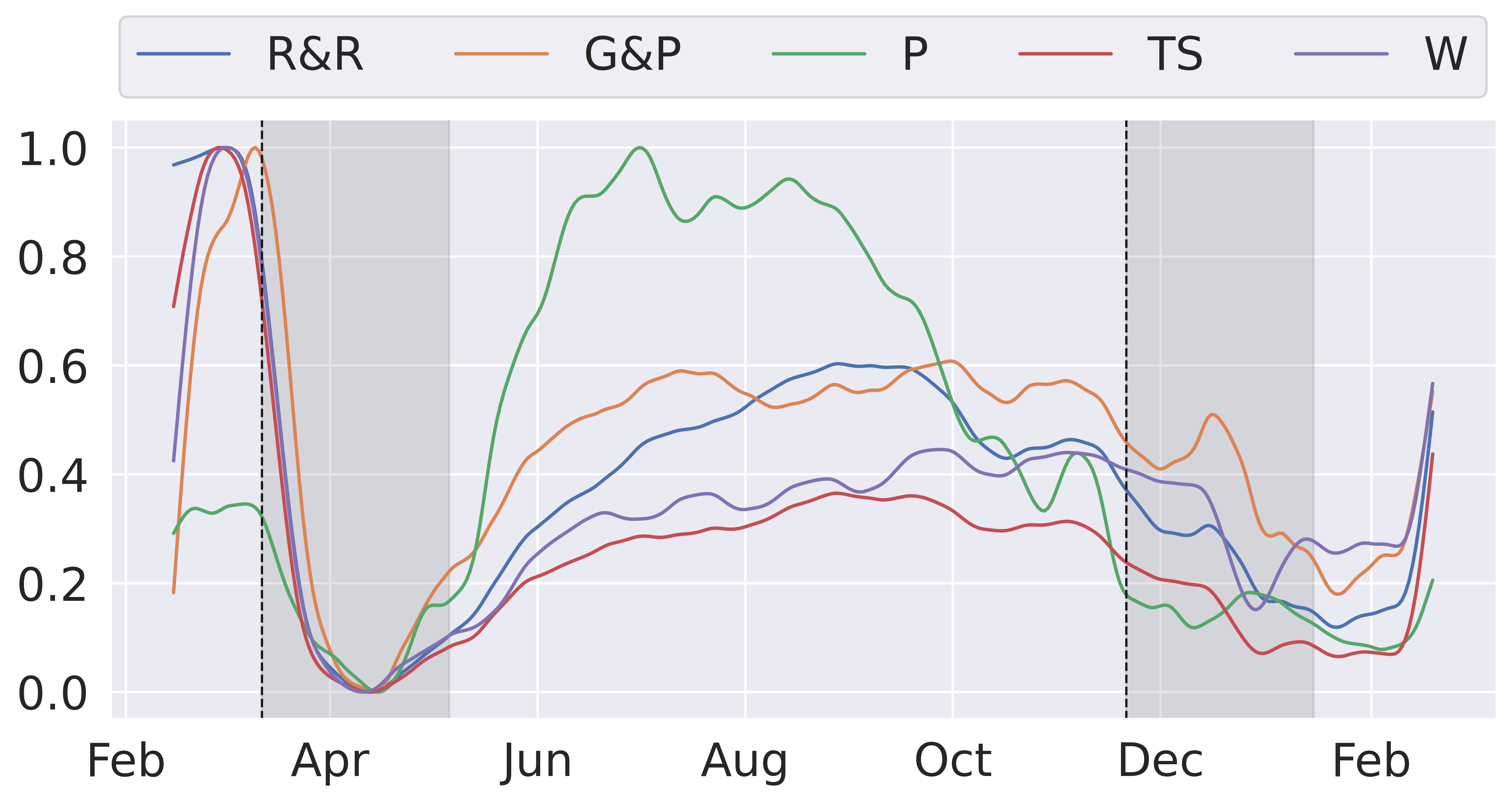}
    \includegraphics[width=\singlefigratio\linewidth, clip=true, trim=100px 30px 100px 0px]{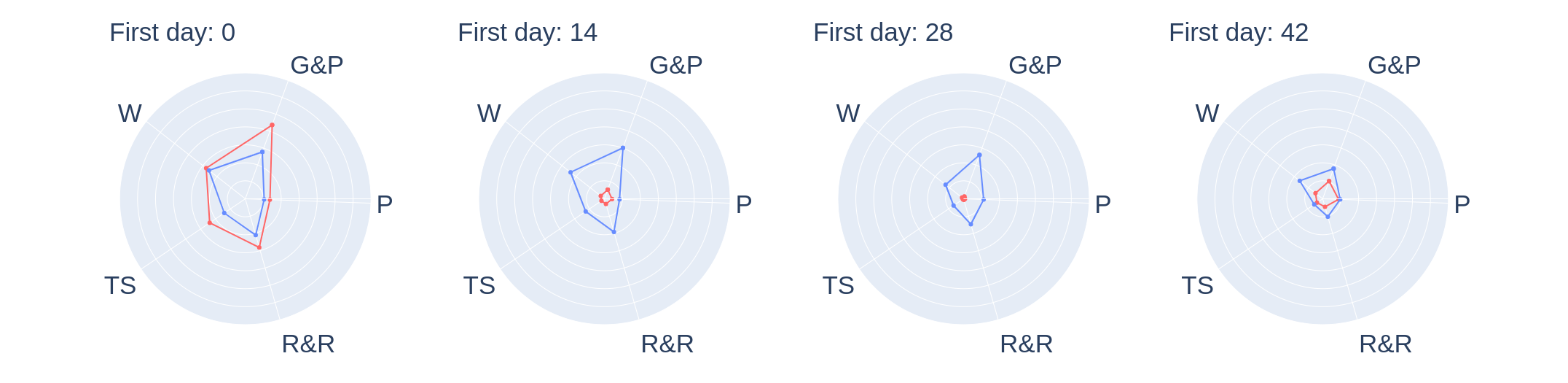}
    	\caption{ Toronto Division}
		\label{fig:toronto5_14}
	\end{subfigure}
	\hfill
    \caption{Temporal mobility analysis for the different localities considered. Left: prepared time series for the different place categories. Initial restriction dates are given as dashed vertical lines, and periods covered by the consecutive windows in both waves are given as shadowed areas. Right: radar charts comparing first~(red) and second~(blue) COVID-19 wave mobility reduction over 14-day consecutive windows. The first day in each window counted as the number of days since the corresponding initial restriction date is given. Charts range from the minimum (0) to the maximum (14) possible AUC value for the period considered. For brevity, place categories in both types of plots are abbreviated. W: \textit{Workspace}; G\&P: \textit{Grocery \& Pharmacy}; P: \textit{Parks}; R \& R: \textit{Retail \& Recreation}; Ts: \textit{Transit stations}.
    %1st row: Lombardia; 2nd row: Île-de-France; 3rd row: Birmingham District; 4th row: Berlin; 5th row: Toronto Division
    }
    \label{FIG:polar_window}
\end{figure*}

\subsection{Lombardia}
\label{sec:lombardia}

Italy’s most wealthy and populous region, Lombardia was by May 2020 the hardest hit region in Europe. Until February 21th, 2021, there were over 577,000 cases and 28,000 deaths in Lombardia according to the Johns Hopkins University, representing almost one fifth and one third of Italian cases and deaths, respectively. During the first wave, Lombardia initially adopted restriction measures regionally on February, 23rd, which were later strengthened and made national by the Italian government. 
Specifically, the second set of restriction measures was decreed on March 8th, prohibiting any kind of mobility apart from certain health or professional needs. A third set of restriction measures was enforced in March 22nd and, among the stricter measures, any sport and physical activity in outdoor spaces was forbidden. Regarding the second wave, the Italian government labelled Lombardia a \textit{red zone} in November 2020, with a rising number of daily new cases. The measures enforced prohibited in and out movement of the city, closed shops, bars, and restaurants, among other measures. 

Figure~\ref{FIG:lombardia70} provides our multi-criteria comparison of the first~(blue) and second~(red) waves of Lombardia, in which per-category data from each 56-day period is 
%We start the analysis of mobility reduction rates in Lombardia considering the 70-day time period as a whole.  for which data is 
aggregated using the area under the curve~(AUC) approach. In more detail, this polar coordinate plot depicts (i)~categories as angles and (ii)~relative mobility aggregated over the given time period as \textit{radii}. More importantly, this multi-dimensional visualization simplifies the Pareto-compliant comparison of the mobility reduction in different waves of a same locality, as mobility reduction from wave $w_1$ dominates mobility reduction from wave $w_2$ iff the polygon given for $w_1$ is contained by the polygon given for $w_2$.
This is the case in this initial analysis given in Figure~\ref{FIG:lombardia70}, where each 56-day period is aggregated as a whole for Lombardia. Indeed, the mobility during the first wave is considerably reduced for all categories in comparison to the mobility during the second wave. This is illustrated in Figure~\ref{fig:lombardia5_14}, where initial restriction dates are given as dashed vertical lines and the 56-day periods from each wave are given as shadowed areas.% replicates the plot given in Figure~\ref{fig:preparation}~(right).
%e observe a higher social distancing adherence during the first wave. 

% A radar chart is employed to depicts AUC aggregation of  \textit{trend} data. In more detail, this polar coordinate plot depicts (i)~categories as angles and (ii)~mobility reduction aggregated over the time period as \textit{radii}. More importantly, this multi-dimensional visualization simplifies the multi-criteria comparison between different waves of a same locality, as wave $w_1$ dominates wave $w_2$ iff the polygon given for $w_1$ is contained by the polygon given for $w_2$. 

%Since 14-day window corresponds to an estimate of epidemiological incubation of the virus, so a multiple-time period analysis is carried out. 
Next, the radar charts given on the right-most part of Figure~\ref{fig:lombardia5_14} depict the multi-objective comparison when each 56-day period is discretized prior to aggregation as 14-day consecutive windows, which we use to assess the temporal evolution of social distance adherence. From left to right, each radar chart depicts a single 14-day window, with the initial day of the window given above the chart, counted as number of days since the corresponding initial restriction date. %The right-most plot gives the prepared time series for Lombardia, with initial restriction dates for each wave given as vertical dashed lines and 56-day periods as shadowed areas~(same as Figure~\ref{fig:preparation}, right).
% considering the first and second waves of Lombardia throughout the consecutive five time periods.
% We assess insights from the multiple-time-period analysis, in which we investigate the effect of the temporal discretization granularity.  Figure~\ref{fig:lombardia5_14} 
% depicts  the  multi-criteria comparison approach considering the first and second waves of Lombardia throughout the consecutive five time periods. 
From the radar charts, we observe that the mobility reduction in the first wave dominates the reduction in the second wave for all windows except for the first~(between 0 and 13 days since each initial restriction). As illustrated in Figure~\ref{fig:lombardia5_14}~(left), this is an effect of the gradual lockdown implementation in Lombardia during the first wave, as previously discussed. Furthermore, having been the first Western country to face the COVID-19 pandemic, Italian society in general did not fully realize the severity of the situation. By contrast, for the second wave the mobility reduction starts even prior to the restriction measures implementation, at a period that coincides with the first official government notice of measures to come.

Finally, we make two remarks concerning scaling effects. 
% Regarding the first wave, 
% concerns how strongly mobility reduction progresses from one window to another. As discussed, this is likely the effect of restriction measures that were incrementally implemented in this locality. In addition, 
As observed in Figure~\ref{fig:lombardia5_14}~(left), the lowest values for all categories across the whole series are observed during the first wave dates that comprise the third and fourth windows. As a result, scaling makes these values very close to zero, reflecting in the very small polygons in the corresponding radar charts.
%the outdoor activity restriction imposed by mid-March. By the third window, the very low values observed f \textit{Parks} mobility, which reaches a very reduced value from the third window on. 
% was affected early and severely by COVID-19. However, the lockdown was implemented in two stages. 
%This may explain why the second wave in the first period (0-14 days) was more effective than the first wave in the same time window. 
% \textcolor{red}{Leo, neste parágrafo a seguir, eu apenas inclui \textit{Grocery \& Pharmacy} pois percebemos tb um aumento junto com \textit{Retail \& Recreation} e a explicação se mantém}.
By contrast, the mobility values for \textit{Grocery \& Pharmacy} during the third and fourth windows of the second wave become very close to maximum, having reached a worrysome pre-pandemic level. Given that the initial restriction date for the second wave in Lombardia is early November, these windows correspond to mid and late December, and this increase is probably explained by the December holiday shopping season. Likewise, all other place categories also see increases in this period, though not to the same extent as for \textit{Grocery \& pharmacy}.
%We believe this is an effect of the rigorous winter that hit Lombardia from the middle of December until the middle of January. 

%Beth
%- primeira onda domina seconda onda em todos os periodos exceto o primeiro (0 a 14 dias). 
%- primeiro periodo da 2a onda: populacao ja vinha com a mobilidade reduzida
%Leo
% na verdade, o problema é na primeira onda: a Lombardia teve um lockdown inicial que foi menos efetivo e duas semanas depois enrijeceu as medidas. 
% na segunda onda, a única variação mais perceptível é o aumento do retail quando chega meio de dezembro, em função das datas comemorativas. 
% dados de parks da segunda onda chegam a zero. Pode ser erro ou efeito do inverno rigoroso (meio de dezembro a meio de janeiro)

%\begin{figure*}[!t]
%    \centering
%    \includegraphics[width=\linewidth, clip=true, trim=50px 100px 30px 60px]{figures/lombardy_polar.png}
%    \caption{Radar charts comparing first~(red) and second~(blue) wave mobility reduction rates in Lombardia over 14-day consecutive windows. The first day in each window counted as the number of days since the corresponding initial restriction date is given above each chart.}
%    \label{fig:lombardia5_14}
%\end{figure*}
\subsection{Île-de-France}
\label{sec:ile_de_france}

The Paris-comprising region of Île-de-France records the highest number of cases and deaths in France to date. As of February 21st, 2021, Île-de-France reports over 70,000 positive cases and 14,000 deaths in total, according to the \textit{Ministère des Solidarités et de la Santé}. During the first wave, Île-de-France adopted a strict lockdown approach, with school suspension and fines for people in the streets without a valid permit.
%, including 60 deaths in 24 hours. 
By the end of October, the government announced France was in the grip of a brutal second wave of the Covid-19 epidemic, with Île-de-France reporting the highest daily infection rates since the beginning of the pandemic. This time, however, schools remained open, and different industries such as construction and cultural were exempted from suspensions. More surprisingly, outdoor spaces such as parks and beaches were not closed, but access to them were restricted to nearby residents and for a limited period of time.

% \textcolor{red}{Leo, pra mim aqui tudo se mantém.}

Figure~\ref{FIG:idf70} depicts the multi-objective comparison of the first and second waves in Île-de-France aggregating each 56-day time period. In contrast to what was observed in Lombardia, the mobility reduction during the first wave dominates the reduction during the second wave, indicating a higher social distancing adherence during the first wave. This is illustrated in~\ref{fig:idf5_14}~(left), and confirmed for all 14-day consecutive windows given % when we assess the temporal evolution of the mobility reduction rates using the  
in Figure~\ref{fig:idf5_14}~(right), 
%depicts  the  multi-criteria comparison approach considering the first and second waves of Île-de-France throughout the consecutive five time periods. Observing this figure, 
where we observe that the mobility reduction during the first wave dominates the reduction during the second wave for all periods considered.
%During that time window, the mobility reduction during the second wave is higher than during the first wave for \textit{Parks}, whereas the opposite happens for the remaining place categories. 
%an effect of a hard winter may explain the reduced  value of \textit{Park} mobility during this window. 

Another observation that is worth notice is the steep decrease in mobility for all categories during the first wave when we progress from the first time window to the remaining.
%, with the exception of the last. 
Since first wave restrictions in France were adopted later than in Italy, society was already more akin to believe in the need for a lockdown, specially with the rise in the number of daily cases and deaths. Finally, Lombardia and Île-de-France are similar as to the mobility behavior during December, depicted in the third and fourth windows of the second wave. In particular, mobility in \textit{Grocery \& pharmacy} reach pre-pandemic levels, and remaining categories also see an increase~(though not to the same extent).
%Though this was also observed in part for Lombardia, the decrease rate in Île-de-France is striking if we consider that no novel restriction measures were imposed at that time at the latter.
%into account the first time window (First Day:0) and the second time window (First Day: 14) for the first wave. It shows a population adherence to the imposed restrictions in the first wave since its pattern is maintained until the forth time window. Only on the last time window, it is possible to see an increase in all categories simultaneously. 

%- primeira onda domina segunda onda em todos os periodos exceto o ultimo (56 a 70 dias) no qual parks apresenta um maior indice na primeira onda em comparacao com a primeira (efeito da estacao do ano??)
%- primeiro periodo da 2a onda: a mobilidade populacao era maior do que na primeira onda (seriedade da populacao?)

%\begin{figure*}[!t]
%    \centering
%    \includegraphics[width=\linewidth, clip=true, trim=50px 100px 30px 60px]{figures/ile_de_france_polar.png}
%    \caption{Radar charts comparing first~(red) and second~(blue) wave mobility reduction rates in Île-de-France over 14-day consecutive windows. The first day in each window counted as the number of days since the corresponding initial restriction date is given above each chart.}
%    \label{fig:idf5_14}
%\end{figure*}
\subsection{Birmingham District}
\label{sec:birmingham}

Birmingham District is the most populated district in England, with Birmingham being the second-largest city in England and the United Kingdom and also the second-largest urban and metropolitan area. Being an international commercial center and an important transport hub, with an economy dominated by the tertiary sector, its population is ethnically mixed encompassing several cultures and religions, which poses a challenge in terms of social distancing adherence. Until February 19th, 2021, there were over 94,000 cases and nearly 2,500 deaths in Birmingham city alone according to the Gov.uk COVID-19 dashboard. During both waves, mobility restriction measures were milder for Birminghan than for the remaining European localities, as previously discussed. Furthermore, the second lockdown started on November 5th and lasted only until December 2nd, having been relaxed for the December holiday shopping season.%, it is possible to see an increase in the \textit{Retail} category.

% \textcolor{red}{Leo, pra mim aqui tudo se mantém.}

Figure~\ref{FIG:birmingham70} depicts the multi-objective comparison between the first and second waves in Birmingham District taking into account the whole 56-day time period for each wave.
% illustrates the  multi-criteria comparison approach considering the first and second waves of Birmingham District taking into account the single 56-day time period. 
Similarly to the analysis of Île-de-France, the mobility reduction for the first wave dominates the mobility reduction for the second wave. However, the differences in mobility reduction for \textit{Parks} is very small, an effect of not having applied restrictive measures regarding this category during either COVID-19 wave. These insights are illustrated in Figure~\ref{fig:birmingham5_14}~(left), though we remark that mobility in \textit{Parks} during the first wave is not as constant as during the second wave.
%Considering the single time  window analysis, the first wave dominates the second wave. 
% The mobility values of all other categories in the second wave are higher than in the first wave. It is somehow expected, since the nurseries are open, support bubbles are allowed, and there are allowances made for care. The residents can do more now than they were allowed in March last year. 

%  Observe that \textit{Parks} is the category responsible for this incomparability: \textit{Parks} in the second wave is smaller than in the first wave. Once again it may be related to the weather with temperatures struggling above freezing. In December 2020 and January 2021, there were several days days of raw, chilly weather and snow in the city.

The 14-day consecutive window analysis given in Figure~\ref{fig:birmingham5_14}~(right) confirms this temporal effect on \textit{Parks} mobility. In particular, only for the second window the mobility reduction during the first wave is greater than during the second. For all other windows, \textit{Parks} renders mobility reduction during the different waves incomparable. Another important remark contrasts with previously discussed European localities. Concerning December mobility, Birmingham district does not present pre-pandemic mobility levels for \textit{Grocery \& pharmacy}, though all categories do see an increase in mobility. \textit{Parks} is again the exception, for which the decrease observed is explained by the heavy snowstorm that hit Birmingham in early December, naturally restricting outdoor activities.
\subsection{Berlin}
\label{sec:berlin}

Berlin, Germany's capital and largest city, is also the most populous city of the European Union, according to population within city limits. Based on the most recent figures, as of February 21st, 2021, Berlin had recorded the highest number of COVID-19 cases in Germany, with nearly 127,000 cases and over 2,700 deaths, according to Johns Hopkins University.
During the first wave, Berlin was not among the more preemptive German regions in terms of promoting social distancing measures, though it followed the measures enforced nationally. Nonetheless, Germany did not adopt a mandatory stay-at-home measure like Île-de-France. For the second wave, Berlin initially stated a semi-curfew and made masks mandatory by mid-October. Given how mild these measures were, we consider as initial restriction date the national light lockdown enforced in early November. In particular, not all measures employed during the first wave were applied for the light lockdown, e.g. schools were kept open. The restriction rules were strengthened by mid-December, including school suspension.
%governments of European countries  imposed a set of restrictions to control the spread of  Covid-19 infections over Christmas, but 
% Germany, in special Berlin, is under tighter rules than other localities since the beginning of November 2020 since coronavirus reaches record levels. The second lockdown, imposing as  local measures in November 2nd, has turned global in Germany since December 16th and it is expected to last until March 2021. 

%OBS: 2o lockdown (local) 02 de Novembro 2020, nacional desde 16 dezembro, ainda em lockdow (espera acabar em Marco)

% - whole period: incomparable due to parks
% - temporal evolution: 
% -- first window: similar, though incomparable due to G&P
% -- second and third: first is more effective than second, though incomparable due to Parks
% -- fourth: second dominates first, as a result of the revised restriction measures

% \textcolor{red}{Leo, pra mim aqui tudo se mantém.}

Figure~\ref{FIG:berlin70} depicts the multi-objective comparison of the mobility in the first and second waves when we aggregate the 56-day period for each wave. Differently from the previously discussed European localities, the mobility reduction rates from both waves are considered incomparable. 
% illustrates the  multi-criteria comparison approach considering the first and second waves of Berlin taking into account the single 56-day time period. Considering this single time analysis, the first and second waves are incomparable.
Mobility for \textit{Parks} was decisive for this, being much higher during the first wave than during the second wave. This is illustrated in Figure~\ref{fig:berlin5_14}~(left), in which mobility reduction for \textit{Parks} during the second wave lasted for a very brief period, having even reached around May a level higher than before the pandemics.
%During the first wave of COVID-19, Germany was acclaimed worldwide for its efforts to combat the coronavirus. Excluding \textit{Parks}, we can observe lower values for all categories in the first wave than in the second one. 

%
The temporal evolution of \textit{Parks} is further detailed in Figure~\ref{fig:berlin5_14}~(right). While for the second wave mobility for this category continually decreased, for the first wave it was only reduced during the first window. From a multi-objective perspective, mobility reduction from both waves are incomparable for all but the last window. Indeed, Figure~\ref{fig:berlin5_14}~(left) confirms the effectiveness of the stricter measures employed by end of December, in contrast to the increased mobility from the holiday shopping season. Regarding the first window comparison given in Figure~\ref{fig:berlin5_14}~(right), mobility reduction from both waves are very similar for most categories, and are considered incomparable because of \textit{Grocery \& pharmacy}. When we examine Figure~\ref{fig:berlin5_14}~(left), we notice that mobility for this category indeed never reduced to the same extent as for other categories nor European localities previously discussed.
\subsection{Toronto Division}
\label{sec:toronto}

Toronto is the largest city in Canada and is considered a world leader in areas such as business, finance, technology, entertainment and culture. Toronto has a large population of immigrants from all over the world, being a multicultural city where enforcing social distancing should be a challenge. As of February 20th 2021, the city of Toronto reports nearly 95,000 cases of COVID-19 and over 2,600 deaths in total according to the Ontario Ministry of Health. During the first wave, Toronto announced (i)~schools would not resume after break in mid-March; (ii)~shutting of non-essential services on the following week, and; (iii)~shutting of outdoor spaces at the end of that month. By the end of November 2020, a variety of wide-scale public health measures, including business and organizational closings, were put into effect in Toronto as a last resort to control the COVID-19 spread. In December 26th, the Ontario government announced a province-wide shutdown applying stricter restriction measures such as closure of schools after the Christmas holiday. 

% \textcolor{red}{Leo, pra mim aqui tudo se mantém.}

Figure~\ref{FIG:toronto70} depicts the multi-objective comparison of the mobility reduction rates during the first and second waves in Toronto Division when aggregating the 56-day period for each wave. Similarly to most European localities, the mobility observed in the first wave is lower than the mobility observed in the second wave for all place categories simultaneously. However, the temporal evolution assessment given in 
% illustrates the  multi-criteria comparison approach considering the first and second waves of Toronto Division taking into account the single 56-day time period. As seen in Lombardia and île-de-France, considering the single time period, the first wave dominates the second wave indicating a higher social distancing adherence during the first wave than the second wave in Toronto. 
Figure~\ref{fig:toronto5_14} shows that the mobility reduction during the second wave was stronger than during the first for the initial time window considered. This results is similar to what was observed in Lombardia, where we observe that the decrease in mobility during the first wave is progressive. This is likely the effect of the incrementally announced restrictive measures, as discussed. Nonetheless, the reductions observed bring mobility to a very low level, which is sustained almost until the end of the first wave. This pattern is reflected on the radar charts given for the remaining time windows~(second to fourth). 

At the same time, we observe that mobility during the second wave decreases gradually from the second to the fourth window considered. This is likely an effect of the stricter measures adopted by late December, which correspond to the third time window given. In fact, for the last time window the mobility in both waves is quite similar, even if not incomparable. More importantly, Toronto Division is the only locality we assess where mobility in December did not increase, with the exception of \textit{Grocery \& pharmacy}. In fact, mobility for most categories often decreased, which could be explained by the recency of restriction measures, implemented at the end of November.
% depicts  the  multi-criteria comparison approach considering the first and second waves of Toronto Division throughout the consecutive four time periods.
%In the first time window (First Day: 0), the second wave dominates the first one. The situation is inverted for the next three time windows: first wave dominates the second one. Also notice the difference between the categories in those windows are high. And, during the last time window (First Day: 56), the waves are incomparable. 
%There was a considerable reduction between first and second window for \textit{Transit Stations} during the second wave. Furthermore, \textit{Transit Stations} reaches zero during second wave from the second time window onward. 

%OBS: 
%Janelas: 
%(i) janela 0: segundo lockdown melhor que primeiro (diferenca pequena)
%(ii) janelas 1, 2 e 3: primeiro lockdown melhor que segundo com diferenca bem pequena para janela 3
%(iii) janela 4: incomparaveis (parks no primeiro lockdown eh muito alto - clima?)

%\begin{figure*}[!t]
%    \centering
%    \includegraphics[width=\linewidth, clip=true, trim=50px 100px 30px 60px]{figures/toronto_polar.png}
%    \caption{Radar charts comparing first~(red) and second~(blue) wave mobility reduction rates in Toronto Division over 14-day consecutive windows. The first day in each window counted as the number of days since the corresponding initial restriction date is given above each chart.}
%    \label{fig:toronto5_14}
%\end{figure*}

\section{Conclusions}
\label{sec:conclusions}

Since the emergence in China back in December 2019 of SARS-CoV-2, the coronavirus responsible for COVID-19, the world has been facing a severe global health crisis. 
%Numerous new cases quickly appeared across all continents,
% in Asian countries, followed by European, American, and  and in the other continents, 
%leading the World Health Organization to declare a public health emergency of international concern on January 30th, 2020. 
Due to the insufficient information on the transmission patterns and the lack of vaccines and specific pharmaceutical treatment alternatives, non-pharmaceutical interventions such as social distancing play an important role for COVID-19 control.  
% Social distancing aims not only to control the spread of the virus but also to avoid the collapse of healthcare system.  
Several countries have implemented a series of social distancing measures such as closing schools, prohibiting mass gatherings, restricting travel, and even enforcing total lockdown to reduce virus transmission. These measures have been introduced gradually and in differing ways, to a greater or lesser extent, locally and globally in the different countries.

In this work, using Google community mobility reports (CMR) data for different localities and a multi-objective time series analysis of mobility reduction, we have compared social distancing in the first and second COVID-19 waves in a Pareto-compliant way. The analysis discussed comprised 56 days since the initial restriction date in each wave, and was further detailed using 14-day consecutive windows to assess for temporal evolution of mobility reduction. 
% has been carried out in a single time period comprising 56 days and in a multiple time period comprising four consecutive 14 days.
Furthermore, the localities assessed represent severely affected regions and/or Europe and America. Though we did not compare localities directly, we have observed how often (i)~mobility reduction during the first wave was stronger than during the second wave; (ii)~the increase in mobility from all categories during the December holiday shopping season, and; (iii)~the contrasting results for \textit{Parks} and \textit{Grocery \& pharmacy}.

Even if results observed are alarming, reduced mobility observed close to the end of the second period assessed sheds hope that more recent restriction measures could help prevent an even stronger disaster. More importantly, it prompts other localities where a second COVID-19 wave is still starting to take strong action from the beginning.

%\section*{Acknowledgment}

%The preferred spelling of the word ``acknowledgment'' in America is without 
%an ``e'' after the ``g''. Avoid the stilted expression ``one of us (R. B. 
%G.) thanks $\ldots$''. Instead, try ``R. B. G. thanks$\ldots$''. Put sponsor 
%acknowledgments in the unnumbered footnote on the first page.

\bibliographystyle{IEEEtran}
\bibliography{refs}

\end{document}